# Planar bridging an Atomically Precise Surface Trench with a Single Molecular Wire on an Au(111) Surface

Umamahesh Thupakula,[1] Xavier Bouju,[1] Jesús Castro-Esteban,[2] Erik Dujardina,[1,3] Diego Peña,[3] and Christian Joachim[1,4]

1 Centre d'Élaboration de Matériaux et d'Études Structurales (CEMES), Centre National de la Recherche Scientifique (CNRS), Université de Toulouse, 29 Rue J. Marvig, BP 94347, 31055 Toulouse Cedex, France.
2 Centro Singular de Investigación en Química Biolóxica e Materiais Moleculares (CiQUS) and Departamento de Química Orgánica, Universidade de Santiago de Compostela, 15782-Santiago de Compostela, Spain.
3 Laboratoire Interdisciplinaire Carnot de Bourgogne, CNRS UMR 6303, Université de Bourgogne Franche-Comté, 9 Av. A. Savary, 21078 Dijon, France.
4 International Center for Materials Nanoarchitectonics (WPI-MANA), National Institute for Material Sciences (NIMS), 1-1 Namiki, Tsukuba, Ibaraki 305-0044, Japan.

## Abstract

In a bridge configuration, a single graphene nanoribbon (GNR) is positioned with a picometer precision over a trench in between two monoatomic steps on an Au(111) surface. This GNR molecular wire adopts a deformed conformation towards the down terrace in between the two contact step edges. Using differential conductance d$I$/d$V$ mapping from a low-temperature scanning tunneling microscope, it is demonstrated how the electronic delocalization along GNR is cut at each contact by its down curvature. It points out the need to bring conductive nanocontacts backside of the support for preserving the front side GNR planar conformation.

## Graphical abstract

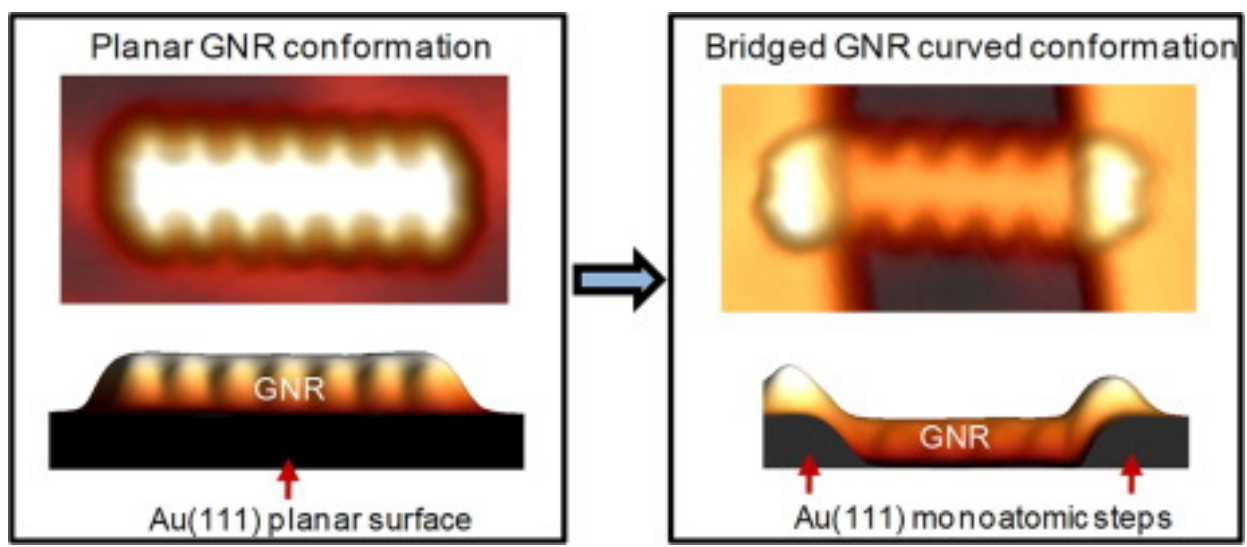

# 1. Introduction

For single molecule electronics where the complete electronic circuit is supposed to be embedded inside one single molecule [1], the planar interconnection of a given molecular wire end to a planar atomically defined metallic nanopads with a picometer precision is a difficult experimental problem [2]. It is also very challenging in terms of understanding the quantum contact conductance between the end of the wire and the surface of the corresponding nanopad [3]. The first vertical contact of a single and always the same molecule with a 20 pm precision [4] and later on a single atom [5] was obtained using the metallic tip of an ultra-high vacuum scanning tunneling microscope (UHV-STM) as a first electrode, while the second electrode was the supporting metallic surface prepared with at atomic precision in the UHV.

To reach a planar configuration with the same pm precision and, after the tentative using of electron beam nanolithography [6], pseudo-planar configurations have been explored with a single on-purpose designed Lander molecule at a mono-atomic step edge of a metal surface [7]. Different step-edge chemical composition and configuration have also been attempted using, for example, ultra-thin insulator nano-islands [8-10]. Short atomic wires have been constructed atom-by-atom on a large gap nanomaterial deposited on a metal surface to demonstrate how such atomic scale connection can be implemented locally [11]. On-surface polymerization was also attempted to reach the single molecule to be wired [12]. Returning to the vertical UHV-STM configuration, the pulling technique of a molecular wire by the STM tip open the path to measure the conductance of a single armchair graphene nanoribbon (GNR) as a function of its length [13,14]. The supporting atomically prepared metallic surface was again the second electrode and the molecular wire maintained in a planar configuration on the surface at least at one end [15].

In this Letter, we report how a conjugated single molecular wire can be positioned in a bridge like configuration over a nanometer scale in width trench in between two monoatomic steps on an Au(111) surface. This configuration is reproducing the conformation of a molecular wire ideally contacted in a planar configuration and with a pm precision between two atomically well-defined metallic contacting nanopads [2]. We have selected the Au(111) surface because (1) on-surface synthesis of long molecular wires is now well mastered [15,16], (2) molecular wires are generally physisorbed on Au(111) as compared for example to the Ag(111) surface [17], and (3) atomically well-defined surface trenches are easily generated on an Au(111) surface [18]. Furthermore, the atomic structure of STM tip apex can be well recovered after nanoindentation.

In section 2, and starting from the diphenyl–10,10′-dibromo-9,9′-bianthracene (DP-DBBA) monomers (laterally extended derivative of DBBA), the on-surface synthesis of periodically modulated with 2 and 6 fused benzene ring wide armchair GNRs, known as 7-13-AGNRs (hereafter simply labeled DP-GNRs), is presented. After the on-surface polymerization and cyclodehydrogenation steps, the experimental conditions to generate mono-atomic in height trenches on the Au(111) surface are discussed. In section 3, low temperature UHV-STM images and molecular manipulations of DP-GNRs having a well selected length are described. With a pm precision and on Au(111), we position a single 7 monomer unit long DP-GNR (7-DP-GNR) in a bridge configuration in-between the two monoatomic step edges of a well selected trench width. Experimental d$I$/d$V$ mappings were recorded to confirm the deformed conformation of the bridging 7-DP-GNR molecular wire. In section 4 and to be more systematic on the passage from a non-deformed DP-GNR planar to a liana bridge like conformation, a triangular shaped trench was used to slide a 9 monomer unit long DP-GNR (9-DP-GNR) step-by-step along the trench. A configuration is reached where one end of the 9-DP-GNR recovers its planar surface conformation on the Au(111) surface while its other end remains contacted atop at the trench step edge. Finally, in the discussion section, the liana bridge distorted conformation is analyzed using molecular mechanics calculations while comparing experimental and calculated constant current scans. It is shown that even in this very elementary bridge like configuration, an exact planar contact conformation of the molecular wire bridging two minimal in height metallic nanopads (relative to the supporting surface) is not possible. While contacted with a pm precision, this points out the urgent need for a *Via* like technology [19] pushed towards the atomic scale to preserve the exact planarity of the molecular wire on its supporting surface.

## 2. Experimental section

On the Au(111) surface, we have performed the on-surface synthesis of the DP-GNRs starting from DP-DBBA monomers (Fig.1a). These molecular wires were selected because their lateral phenyl groups are supposed to stabilize the end contact and also because its ground to first excited states energy separation (approximately equals to its HOMO-LUMO gap) is believed to be smaller than the well-known 7–GNRs resulting also from the on-surface polymerization of the DBBA monomers [16]. On-surface polymerization of the DP-DBBA monomers on an Au(111) surface was already described in the literature but a very high density of oligomers was reported [20]. To drastically reduce this density, we have limited the initial monomer precursor concentration on Au(111) surface and further carried out a two-step annealing approach.

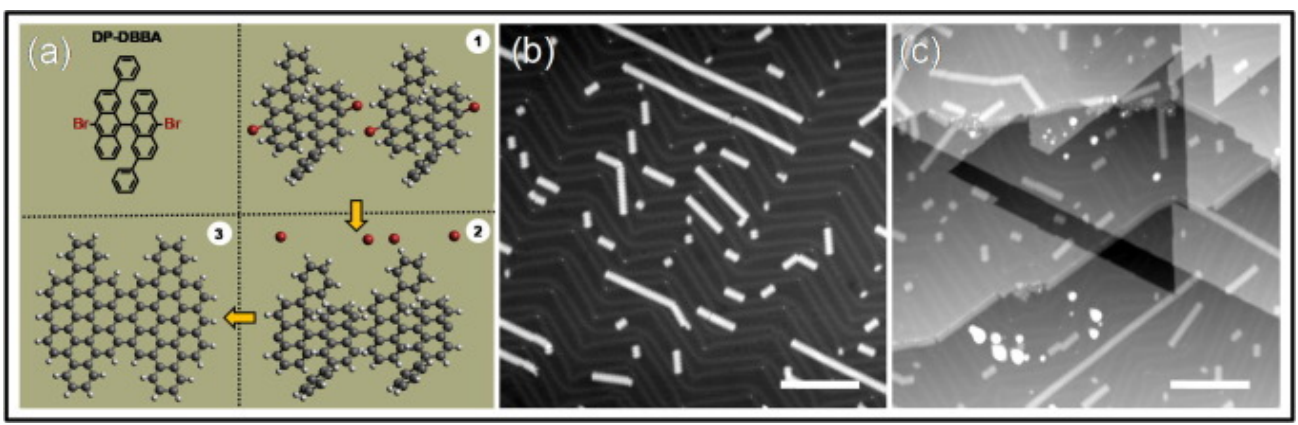

*Fig. 1. On-surface synthesis of DP-GNRs and generation of monoatomic height trenches on the Au(111) surface. (a) Schematic illustration showing the on-surface synthesis of DP-GNRs with the basic DP-DBBA monomer precursor (top left), the ball model of two such DP-DBBA monomers (top right frame labeled 1), the polymerization reaction (bottom right frame labeled 2) and finally the cyclodehydrogenation reaction (bottom left frame labeled 3). (b) Large scale constant-current STM image with different in length DP-GNRs on the Au(111) surface. (c) The STM image of a linear and the triangular sharp terraces and trenches reconstruction after a gentle STM tip nano-indentation process on the Au(111) surface. Scale bars 20 nm. STM set-parameters are +100 mV/100 pA.*

### 2.1. Au(111) surface preparation

Single crystal (111) oriented Au surface (*Mateck* GmbH) was used as the atomically flat support for the on-surface synthesis and UHV-LT STM single molecule molecular manipulation experiments. Initially, the Au(111) surface was cleaned with repeated cycles of Ar+ ion sputtering (for 10 minutes at ~1.0 keV and ~$10^{-6}$ mbar) and subsequent annealing under UHV conditions at about 750 K for 1h. The quality of the Au(111) surface was checked for its cleanliness using any one STM head of our LT-UHV 4-STM instrument operating at a sample stage temperatures of ~4.5 K [21]. A typical herringbone surface topography reconstruction and the characteristic Au(111) surface state were observed, indicative of atomically flat and clean Au(111) surface as well as the chemical composition of the STM tip apex. All the samples were prepared in UHV conditions, transferred and characterized in situ in our LT-UHV 4-STM instrument (base pressure <3×$10^{-11}$ mbar, *Scienta Omicron*). A lock-in detection technique was used for recording the DP-GNR d*I*/d*V* maps in open feedback loop conditions ($V_{mod}$ = 16 mV at ~430 Hz). STM and STS data were acquired with electrochemically etched Pt/Ir (80/20) tips.

## 2.2. DP-DBBA polymerization and cyclodehydrogenation on Au(111)

The DP-DBBA monomers were thermally evaporated from a quartz crucible of a Knudsen-cell thermal evaporator (*Kentax* triple cell system) at 160 °C for 10 minutes onto the clean Au(111) surface maintained at room temperature in our UHV preparation chamber. Starting from a 1.0 ML of DP-DBBA coverage, a two-step on-surface synthesis was performed (as illustrated in Fig.1a) by annealing the sample first at ~200 °C and then at 400 °C for 10 minutes at each temperature. It results first the formation of long DP-DBBA oligomer chains via an Ullmann type coupling reaction (at the bromine sites) and then the DP-GNRs via cyclodehydrogenation reaction, respectively. Figure1b presents a typical 100 nm × 100 nm STM image recorded in our LT-UHV 4-STM instrument. This on-surface synthesis strategy leads to a very low density of DP-GNRs with a length distribution ranging from the DP-DBBA monomer to ~80 nm long DP-GNRs.

## 2.3. Creating the mono atomic high trenches on Au(111) after the DP-GNR synthesis

Directly after the DP-GNRs synthesis, a gentle STM tip apex indentation was performed maintaining the Au(111) surface on the LT-UHV 4-STM stage at liquid helium temperatures. For this purpose, the STM tip of the selected STM head was initially scanned in a constant current mode (set-point: $V$ = +0.1 V and $I$ = 100 pA) to identify a clean Au(111) surface region of at least 20 nm × 20 nm with no molecules. This avoids a possible pickup of DP-GNRs by the STM tip apex during the indentation process. Next, the STM tip apex was positioned at the center of this clean Au(111) region and the nano-indentation performed using the STM atom manipulation mode with a targeted 200 nA current at 4 mV bias voltage. This is equivalent to a 20 kΩ STM tunnel junction resistance, just below the quantum contact resistance regime of our 4-STM setup. At these conditions, approaching for few seconds the STM tip apex down to a minimum of 5 nm in '$\Delta z$' relative to the Au(111) surface and sliding the tip <1 nm in the lateral *x-y* plane generates well defined dislocation emerging at the Au(111) surface [18]. This lateral tip apex displacement step is required to complete the nanoindentation procedure [18] in a reproducible manner to obtain long and well-defined atomic trenches as presented in Fig.1c. Then, the STM tip apex was reshaped at a different free of molecule location to re-define a metallic and artifact free tip apex (accessed by probing the Au(111) surface state). Thanks to this clean and gentle indentation procedure, dislocations emerge randomly on an Au(111) surface region of approximately 500 nm × 500 nm around the tip apex impact with a large distribution of mono atomic high triangular steps and linear trenches of different widths. After this indentation step, a systematic exploration of the resulting Au(111) reveals that the surface deformations take place without affecting the quality of DP-GNRs chemical structures. It allows selecting the DP-GNRs with the precise lengths suitable for bridging linear and triangular mono-atomic in height trenches (Fig.1c).

## 2.4. Theory simulation details

Semi-empirical molecular mechanics calculations were performed using the ASED+ code including van der Waals interactions. It was parametrized using DFT calculations for small unit cell surface [22]. Constant current STM scans were calculated using the Elastic Scattering Quantum Chemistry semi-empirical (ESQC-EHMO) code [23] using the generalized Landauer-Buttiker formula to calculate the current intensity avoiding the Bardeen approximation and leading to a virtual STM like procedure, which is able to calculate the $\Delta z$ constant current scan corrugation. A double zeta basis set was used at the tip apex end atom to reproduce the Au(111) work function.

## 3. Bridging the trench

After having located a linear trench with its two-parallel mono-atomic step edges on the Au(111) surface, a length compatible DP-GNR was selected and manipulated by STM in a bridge like configuration across the trench (Fig.2a). The width of the Fig.2a trench is ~3.70±0.02 nm down the trench (Fig.2b and see supporting information Figure S1). Considering a planar non-deformed DP-GNR bridging this trench, a DP-GNR with seven monomer moieties was selected corresponding to a ~5.81 nm distance between edge-C to edge-C of the two terminal anthracene units of a 7-DP-GNR. The corresponding experimental LT-UHV STM line profiles extracted along a given flat 7-DP-GNR and across the selected linear trench (as presented in Fig2a and 2b) confirm that a 7-DP-GNR is perfectly suitable for a bridge like configuration with one monomer unit physisorbed atop each mono-atomic step edge.

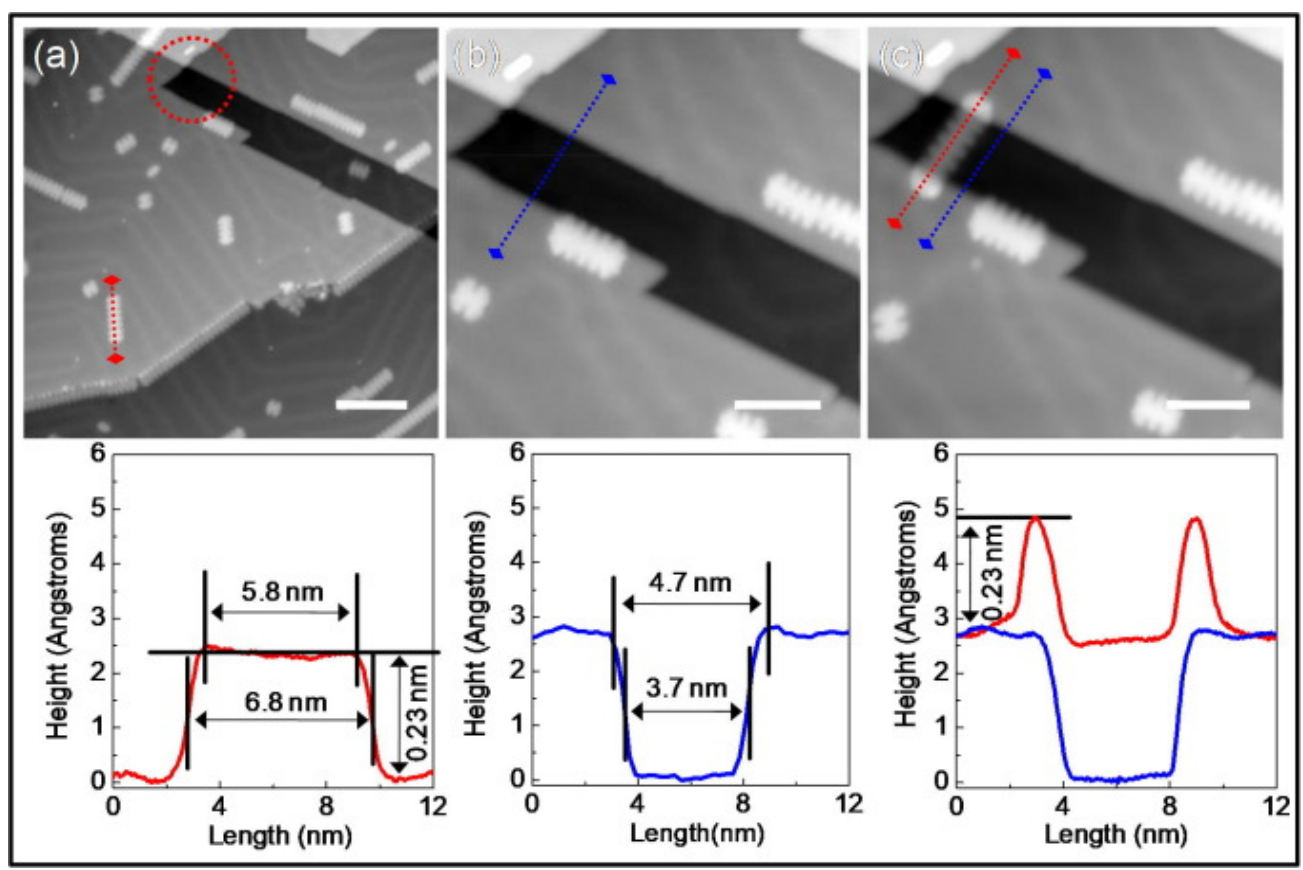

*Fig. 2. (a) A large scale STM image showing the linear monoatomic height trench fabricated on Au(111) (marked with a red dotted circle) and a 7-DP-GNR (under the red dotted line down left). Corresponding STM images before (b) and after (c) the 7-DP-GNR manipulation in a bridge like conformation on the identified linear trench in (a). The experimental Δz line profiles recorded across a flat 7-DP-GNR (bottom left (a)), the linear trench (bottom middle (b)) and in the bridged conformation (bottom right (c)) are indicated with the corresponding colored dotted line on the STM images. Scale bars are 10 nm in (a) and 5 nm in (b,c). STM set-parameters are +100 mV/100 pA. The Δz 7-DP-GNR STM corrugation (red) is the same in (a) and (c) while taking the top terrace as a reference for (c).*

Before its STM molecular manipulation [24], the selected 7-DP-GNR was STM imaged and its d$I$/d$V$ spectroscopy and maps were also recorded for reference as presented in Fig.3a and 3c (see also supporting information Figure S2). For STM molecular manipulation and to obtain the bridge configuration, the STM tunnel junction resistance was reduced from 1 GΩ to 1 MΩ. Starting from the Fig.3a planar physisorbed conformation, the 7-DP-GNR step-by-step constant current molecular manipulation sequence is presented in the supporting information Figure S3 leading to the bridged configuration. The same molecular manipulation conditions were also employed for the section 4 experiments where this time a DP-GNR with 9 monomer units (9-DP-GNR) was positioned and then pushed step-by-step away from a triangle like trench (as discussed later in section 4 and Fig.4 below). After molecular manipulations, the STM images and line profiles of the 7-DP-GNR in its bridged configuration confirm its exact matching with the selected linear trench width (Fig.3b, 3d and 3f). The corresponding optimized molecular mechanics conformation bridging the trench is also presented in Fig.3f (see discussion below and supporting information Figure S6).

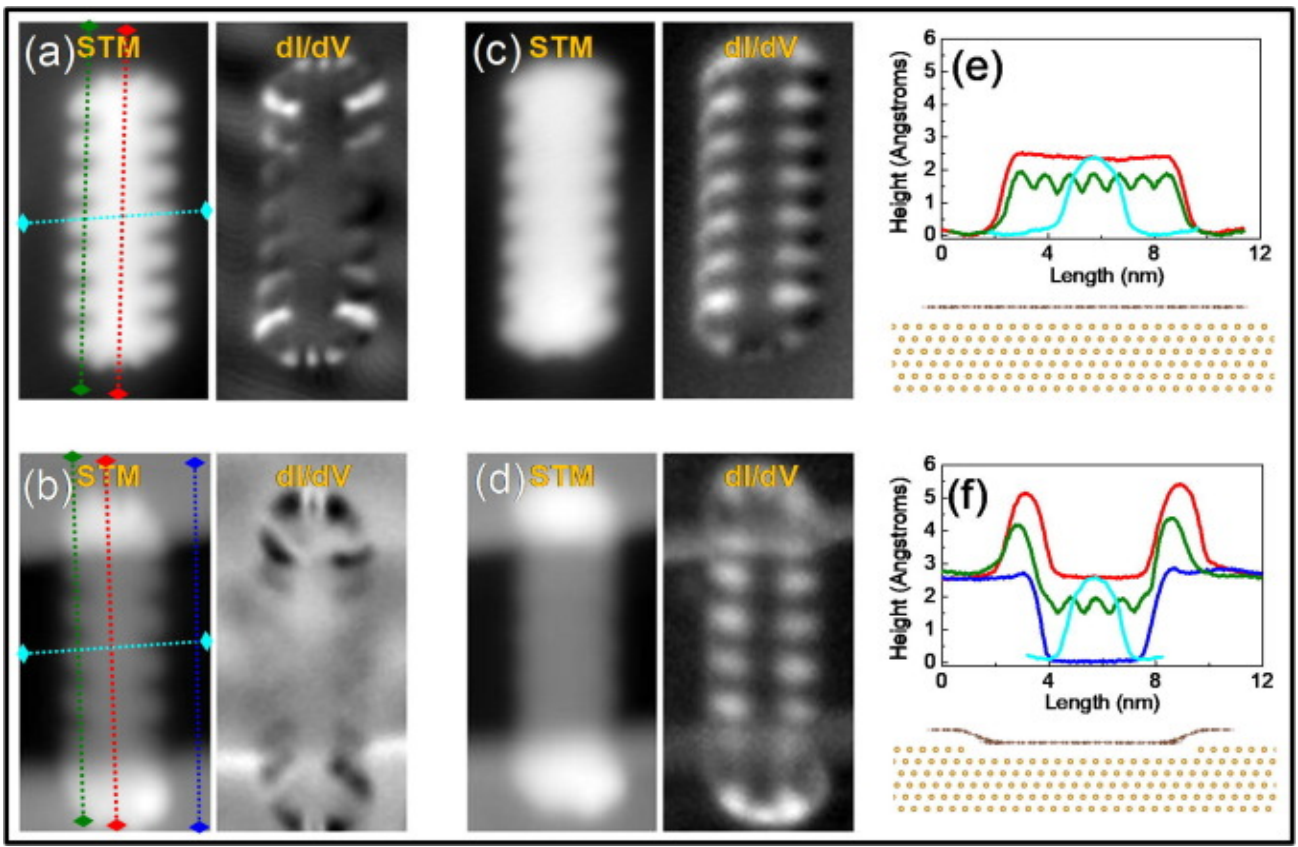

*Fig. 3. **Comparison of electronic differential conductance**. The experimental constant current STM images (left) and corresponding dI/dV maps (right) recorded on a 7-DP-GNR before (a,c) and after (b,d) bridged conformation respectively at +100 mV/100 pA (a,b) and at +2V/100 pA (c,d). All scan sizes are 4 nm × 8 nm. The corresponding line profiles indicated with dotted lines in (a) and (b) are presented in (e) and (f) with the same color, respectively. The theoretical optimized conformations of the 7-DP-GNR before and after bridging are presented below (e) and (f), respectively.*

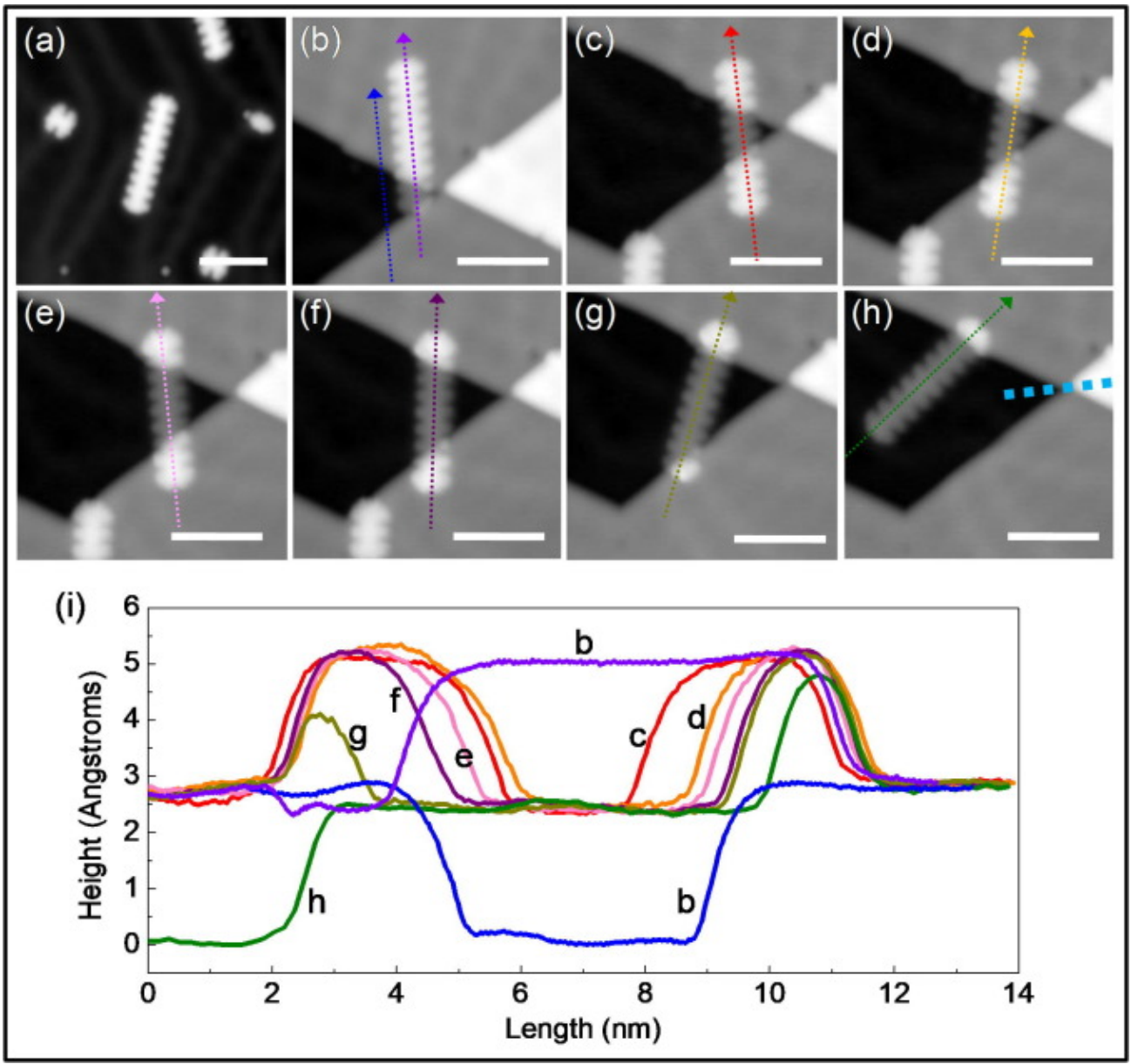

*Fig. 4. **Manipulation of 9-DP-GNR across a triangular trench.** (a-h) the STM image sequence showcasing the 9-DP-GNR, the triangular trench and the step-by-step manipulation sequence of 9-DP-GNR across the triangular trench. (i) Corresponding height profile change from (b) to (h) along the 9-DP-GNR at each manipulation step showing how the 9-DP-GNR is conforming to a down planar configuration already from the very first step (b) at the narrowest trench. The line profiles are indicated with colored dotted arrow directions on the STM images. Scale bars are 5 nm. STM set-parameters are +100 mV/100 pA.*

The d$I$/d$V$ maps were recorded at bias voltages of +0.1 V (Fig.3b) and +2.0 V (Fig.d) where we have identified the first two tunneling electronic resonances for DP-GNRs. The d$I$/d$V$ map recorded at +0.1 V reveals a bright contrast for the planar GNR (Fig.3a) whereas a dark contrast is observed for the bridged 7-DP-GNR (Fig.3b). This bright contrast for the planar 7-DP-GNR comes from the known edge electronic states of a GNR [15] that appear at the Fermi level. In a bridging configuration, the 7-DP-GNR liana is strongly deformed (Fig.3f) resulting in a down shift in energy of this resonance hence a dark contrast is captured while mapping at +0.1 V.

Comparing now the d$I$/d$V$ maps recorded at +2.0 V for the planar (Fig.3c) and bridged (Fig.3d) configurations reveals a small contrast change in the middle of the molecular wire corresponding to a central flat conformation. The interest of the +2.0 V resonance is that it corresponds to an electronic state delocalized along the 7-DP-GNR molecular structure with its multi-bay regions

(formed by the periodic modulation of the 7– and 13–carbon atom–wide sections of DP-GNRs, see Fig.1a). The +0.1 V resonance corresponds to edge molecular states dying off exponentially along the molecular wire after the trench step edge. At +2.0 V, the number of differential conductance bumps and multi-bay nodes along the 7-DP-GNR are the same in both configurations. In first approximation, the number of bumps correspond to the maximum of differential conductance coming from the LUMO mono-electronic content of this +2.0 V tunneling resonance. Identical to a non-bridged 7-DP-GNR conformation for the central part, this central flatness is confirmed by comparing the experimental transverse light blue scans in Fig.3e and 3f (see also the comparison of experimental *vs* calculated line scans presented in supporting information Figure S4). Exactly at the step edges, the d$I$/d$V$ map is affected by the step edge induced molecular deformation with a lower differential conductance resulting in smaller bump intensity as compared to the planar case. This comes from a local molecular structure deformation at each step edge which smears out the molecular orbital content of the +2.0 V resonance (see supporting information Figure S5). Notice also that there is a minor lateral deformation by comparing both d$I$/d$V$ maps at + 2.0 V (planar and bridged) central. In Fig.3c d$I$/d$V$ map, the lateral semi-circular extensions between the conductance bumps extending laterally outside the 7-DP-GNR central are absent on the Fig.3d d$I$/d$V$ map central. For a flat Au(111) surface configuration, it is characteristic of the Au(111) 2D free electronic waves scattered by the multi-bay along the molecular wire. This phenomenon is absent in the constrained mono-atomic trench where laterally quantified Au(111) standing wave eigen modes are already existing.

## 4. Different configurations along a triangle like trench

In order to follow the progressive planarization of the central part of a bridged DP-GNR, a 9-DP-GNR was selected to be slid step-by-step along a triangular trench as presented in Fig.4. Here, a large portion of this 9-DP-GNR is physisorbed and lays flat at one end on the Au(111) surface. For this experiment, a 9-DP-GNR was chosen instead of a 7-DP-GNR in order to completely electronically decouple the two edge states through the molecular wire [25] since no d$I$/d$V$ tunneling resonance splitting can be measured for the edge states using a DP-GNR longer than the 7-DP-GNR. In Fig.4a, we started with the 9-DP-GNR flat molecular wire, manipulated it towards by STM and then along a well identified and clean triangular trench (Fig.4b). At each step, from Fig.4b to 4h, corresponding constant current scan lines can be extracted for comparison along the 9-DP-GNR. It was not possible along the manipulation sequence as shown in Fig.4 to respect the orthogonality of the 9-DP-GNR relative to the triangle axle (indicated by the dashed cyan line in Fig.4h) because of the [111] surface orientation which is imposing for each manipulation step a registry with the Au(111) surface. While sliding 9-DP-GNR step-by-step along the triangle, the molecular wire deformation identified in Fig.3 already takes place at the narrowest possible in this experiment trench width (Fig.4b). It settles down towards a central planar conformation as soon as the trench width is greater than two DP-DBBA monomer units (Fig.4c). The cumulative longitudinal constant current line scans presented in Fig.4i is indicative of the progressive central planarization of the 9-DP-GNR molecular structure. Particularly, when the trench width matches 2-3 units of the 9-DP-GNR length (Fig.4c and 4d, and see supporting information Figure S6), the line scans extracted at the initial and intermediate steps of the 9-DP-GNR manipulation already demonstrate the flat conformation of the 9-DP-GNR down the trench. It matches with the line profile extracted at the flat end conformation of the 9-DP-GNR (Fig.4h).

## 5. Discussion

In a two nanopads like configuration, the conformation of a molecular wire bridging a trench as presented above depends on the width and depth of the trench. In addition, this conformation is also dictated by the mechanical flexibility of the wire above the trench. We assume here that the adsorption of the ends of the molecular wire at the step edges of Au(111) trench is strong enough for the wire not to slip and fall completely down at the bottom of the trench. Notice that the width of the trench is smaller than the molecule length. As studied here and for a wide enough trench, the molecular wire is settled down at the center of the trench to reach a flat conformation. This is counter balanced by the limited molecular structure flexibility of the molecular wire but favored by the trench step edge slope (for example a [111] facet will help the curvature as compared with a [100] one).

We have performed systematic semi-empirical ASED+ molecular mechanics optimizations [22] to quantify the competition between central attraction towards the surface and curvature limitation of the 7-DP-GNR molecular wire in the adsorption configuration constructed experimentally in Fig.2. The molecular side view models presented in Fig.3 are resulting from those optimizations. Starting from the flat Fig.3e one towards the Fig.3f conformation, all the details of studied configurations are presented in the supporting information Figure S6. We have systematically varied the width and the depth of the trench with always the same 7-DP-GNR length. The balance between van der Waals attractions down the trench and the molecular wire energy increase due to its resulting curvature is presented in Fig.5a. For an Au(111) down surface, the attraction of the down surface starts to operate and to deform the 7-DP-GNR when one DP-DBBA unit is free from the step edges as observed experimentally (see also supporting information Figure S6, trench width 17.59 Å). Then, increasing the trench depth accentuates the 7-DP-GNR edge curvature up to the point where a flat conformation is recovered for a too deep trench. Other chemical composition of the down surface or at the step edge will lead to a different balance. The effect is similar to what was already observed experimentally with a one-step edge molecular wire interactions [7-9].

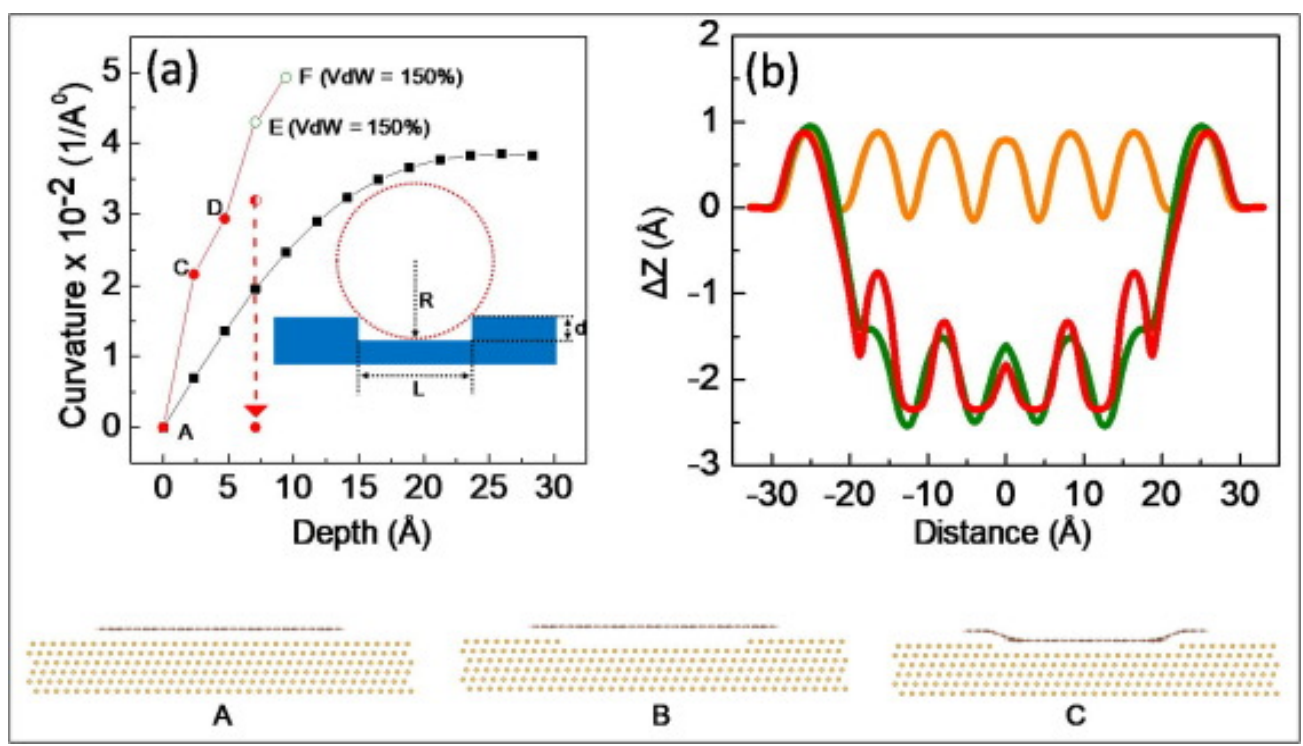

*Fig. 5. (a) Curvature (square data for model rigid osculator and circle data for model 7-DP-GNR) of a bridging 7-DP-GNR as a function of the trench depth. The simple model of the black data is shown as inset: a rigid osculator (dotted red) circle (radius R) is inserted in a trench of length L and depth d (L = 51.93 Å and d is taken as a multiple of 2.36 Å). The curvature 1/R varies as $8d/(L^2+4\times d^2)$. The more accurate calculation in the presence of the van der Waals (vdW) forces exerted by the lower surface on the suspended section the GNR also yields a curvature that initially increases curvature in supported conformation (full red circles) as the trench depth increased. However, when the vdW attraction is too small to pin the GNR at the bottom of the trench, the 7-DP-GNR switches to a flat (null curvature) suspended conformation (red dash arrow). The open green circles are calculated by artificially increasing the vdW attraction*

*forces by 150%. This restores the contact with the lower surface for deeper trenches. (b) The calculated constant current scans along 7-DP-GNR in the optimized (A, orange scan), (C, green scan) and artificially suspended rigid planar bridging (B, red scan) conformations respectively. The contact bump [2] for B is followed by its 3 consecutive contrast maxima along the wire. ESQC-EHMO calculations are with 10 mV/100 pA.*

Constant current line scans have been calculated using ESQC-EHMO [23] along the Fig.3 planar and bridged liana configurations using the optimized ASED+ surface conformations. For clarity, they are presented in Fig.5b for a scan longitudinal and running on the lateral phenyl rings, in the supporting information Figure S4 for transverse scans at a step edge contact and Figure S7 for a scan longitudinal and running at the center of the 7-DP-GNR. Calculated longitudinal scans are identical to the Fig.3 experimental ones. They confirm how the 7-DP-GNR is not adopting a planar rigid bridge like conformation but rather goes down center flat on the Au(111) lower terrace as quickly as one monomer unit per step edge. This phenomenon was also observed for different step edges or surface chemical structures, in particular, at insulating single step edge surfaces [7-9]. Such deformation is very damageable for the contact conductance on a molecular wire. For example, and calculating the longitudinal constant current scan along an artificially maintained planar conformation (Fig.5b, red line), the contact bump at both step edges is followed by a $\Delta z$ corrugation decay of the molecular orbital like bump envelop along the wire [3]. In the optimized curved conformation, both-end contact bumps are down in corrugation (Fig.5b, green line) with now no decay along the molecular wire. As a consequence, and in this curved conformation, the electronic delocalization along the wire is cut by this curvature (compare the experimental Fig.3f and the red rigid Fig.5b calculated scans). This is drastically attenuating the way the metallic surface wave function coming from the two contact step edges through this molecular wire (compare the Fig.5b red and green scans). Notice that the calculated $\Delta z$ constant current corrugation is different from the experimental one because in the semi-empirical mono-electronic ESQC-EHMO approximation, the HOMO-LUMO gap is under-estimated.

In conclusion, between two mono-atomic step edges mimicking a planar contact and using pm precision LT-UHV 4-STM scan lines and differential conductance mapping, we have experimentally demonstrated how a bridged molecular wire based on a GNR is adopting a deformed conformation towards the down terrace between the two step edges even for mono-atomic Au(111) step edges. This is reducing the way the metallic wave function coming from the contact step edges is penetrating through the molecular wire because the electronic delocalization along this wire is suppressed by the curvature of the molecular wire at each contact edge. To keep a good contact conductance [3], the planarity of a bridge between two contacting nanopads must be preserved. Therefore, it is now important to bring contact nanopads from the backside of the sample in a *Via* like technology [19] for preserving the exact planarity of the molecular wire on its top side supporting surface.

**Author contributions**

UT was responsible with CJ for planning and conducting the whole STM experiments XB for performing the calculations, JCE and DP for the synthesis of monomers and ED with UT for the on-surface chemistry procedure.

**Notes**

The authors declare no competing financial interest.

**Declaration of Competing Interest**

The authors declare that they have no known competing financial interests or personal relationships that could have appeared to influence the work reported in this paper.

**Acknowledgement**

This work was supported in part by the World Premier International Center (WPI) Initiative on Nanoarchitectonics (MANA-NIMS), MEXT (Japan), by the European H2020 Marie Curie project "GNR Conductance" (Grant N° 895239) and by a granted access to the HPC resources of CALMIP supercomputing center under the allocation 2021-P0832. In addition, this work was supported by the Spanish Agencia Estatal de Investigación (PID2019-107338RB-C62 and PCI2019-111933-2), Xunta de Galicia (Centro de Investigación de Galicia accreditation 2019–2022, ED431G 2019/03), and the European Regional Development Fund (ERDF).

# REFERENCES

[1] C. Joachim, J.K. Gimzewski, A. Aviram
**Electronics using hybrid- molecular and mono- molecular devices**
Nature, 408 (6812) (2000), pp. 541-548

[2] C. Joachim, D. Martrou, M. Rezeq, C. Troadec, D. Jie, N. Chandrasekhar, S. Gauthier
**Multiple atomic scale solid surface interconnects for atom circuits and molecule logic gates**
J. Phys. Cond. Mat., 22 (2010), Article 084025

[3] S. Stojkovic, C. Joachim, L. Grill, F. Moresco
**The contact conductance on a molecular wire**
Chem. Phys. Lett., 408 (2005), p. 134

[4] C. Joachim, J.K. Gimzewski, R.R. Schlittler, C. Chavy
**Electronic transparence of a single $C_{60}$ molecule**
Phys. Rev. Lett., 74 (1995), p. 2102

[5] A. Yazdani, D.M. Eigler, N.D. Lang
**Off-resonance conduction through atomic wires**
Science, 272 (5270) (1996), pp. 1921-1924

[6] M.S.M. Saifullah, T. Ondarçuhu, D.F. Koltsov, C. Joachim, M. Welland
**A reliable scheme for fabricating sub-5 nm co-planar junction for molecular electronics**
Nanotechnology, 13 (2002), p. 659

[7] F. Moresco, L. Gross, M. Alemani, K.-H. Rieder, H. Tang, A. Gourdon, C. Joachim
**Probing the Probing the different stages in contacting a single molecular wire**
Phys. Rev. Lett., 91 (2003), Article 036601

[8] C. Bombis, F. Ample, L. Lafferentz, H. Yu, S. Hecht, C. Joachim, L. Grill
**Single molecular wires connecting metallic and insulating surface areas**
Angew. Chem. Int. Ed., 48 (2009), p. 9966

[9] P.H. Jacobse, M.J.J. Mangnus, S.J.M. Zevenhuizen, I. Swart
**Mapping the conductance of electronically decoupled graphene nanoribbons**
ACS Nano, 12 (7) (2018), pp. 7048-7056

[10] S. Wang, N. Kharche, E.C. Girão, X. Feng, K. Müllen, V. Meunier, R. Fasel, P. Ruffieux
**Quantum dots in graphene nanoribbons**

Nano Lett., 17 (2017), p. 4277

[11] G.V. Nazin, X.H. Qiu, W. Ho
**Visualization and spectroscopy of a metal-molecule-metal bridge**
Science, 302 (5642) (2003), pp. 77-81

[12] N. Nakuya, Y. Okawa, C. Joachim, M. Aono, T. Nakayama
**Surface creation of nanojunction between fullerene and 1D conductive polymers**
ACS Nano, 8 (2014), p. 12259

[13] L. Lafferentz, F. Ample, H. Yu, S. Hecht, C. Joachim, L. Grill
**The conductance of a single conjugated polymer as a continuous function of its length**
Science, 323 (5918) (2009), pp. 1193-1197

[14] G. Reecht, F. Scheurer, V. Speisser, Y.J. Dappe, F. Mathevet, G. Schull
**Electroluminescence of a polythiophene molecular wire suspended between a metallic surface and the tip of a scanning tunneling microscope**
Phys. Rev. Lett., 112 (2014), Article 047403

[15] M. Koch, F. Ample, C. Joachim, L. Grill
**Voltage-dependent conductance of a single graphene nanoribbon**
Nat. Nanotechnol., 7 (2012), p. 713

[16] J. Cai, P. Ruffieux, R. Jaafar, M. Bieri, T. Braun, S. Blankenburg, M. Muoth, A.P. Seitsonen, M. Saleh, X. Feng, K. Müllen, R. Fasel
**Atomically precise bottom-up fabrication of graphene nanoribbons**
Nature, 466 (7305) (2010), pp. 470-473

[17] K.A. Simonov, A.V. Generalov, A.S. Vinogradov, G.I. Svirskiy, A.A. Cafolla, C. McGuinness, T. Taketsugu, A. Lyalin, N. Mårtensson, A.B. Preobrajenski
**Synthesis of armchair graphene nanoribbons from the 10,10′-dibromo-9,9′-bianthracene molecules on Ag(111): the role of organometallic intermediates**
Sci. Rep., 8 (2018), p. 3506

[18] A. Asenjo, M. Jaafar, E. Carrasco, J.M. Rojo
**Dislocation mechanisms in the first stage of plasticity of nanoindented Au(111) surfaces**
Phys. Rev. B, 73 (2006), Article 075431

[19] D. Sordes, A.Thuaire, P. Reynaud, C. Rauer, J.-M. Hartmann, H. Moriceau, E. Rolland, M. Kolmer, M. Szymonski, C. Durand, C. Joachim, S. Chéramy and X. Baillin, *Nanopackaging of Si(100)H wafer for atomic scale investigation*. In "On-surface Atomic Wires and Logic Gates" Springer Series: Advances in Atom and Single Molecule Machines: Vol. IX, p. 25 (2017) ISBN 978-3-319-51846-6

[20] C. Moreno, M. Vilas-Varela, B. Kretz, A. Garcia-Lekue, M.V. Costache, M. Paradinas, M. Panighel, G. Ceballos, S.O. Valenzuela, D. Peña, A. Mugarza
**Bottom-up synthesis of multifunctional nanoporous graphene**
Science, 360 (6385) (2018), pp. 199-203

[21] J. Yang, D. Sordes, M. Kolmer, D. Martrou, C. Joachim
**Imaging Single atom contact and single Atom manipulation at Low Temperature using the new Scienta Omicron LT-UHV 4 STM**

Eur. Phys. J. Appl. Phys., 73 (2016), p. 10702

[22] F. Ample, C. Joachim
**A semi-empirical study of polyacene molecules adsorbed on a Cu(110) surface**
Surf. Sci., 600 (2006), p. 3243

[23] P. Sautet, C. Joachim
**Interpretation of STM image: Copper-Phthalocyanine on Copper**
Surf. Sci., 271 (1992), p. 387

[24] S.-W. Hla
**Scanning tunneling microscopy single atom/molecule manipulation and its application to nanoscience and technology**
J. Vac. Sci. Technol. B, 23 (4) (2005), p. 1351

[25] J.-P. Launay
**Long-distance intervalence electron transfer**
Chem. Soc. Rev., 30 (2001), p. 386

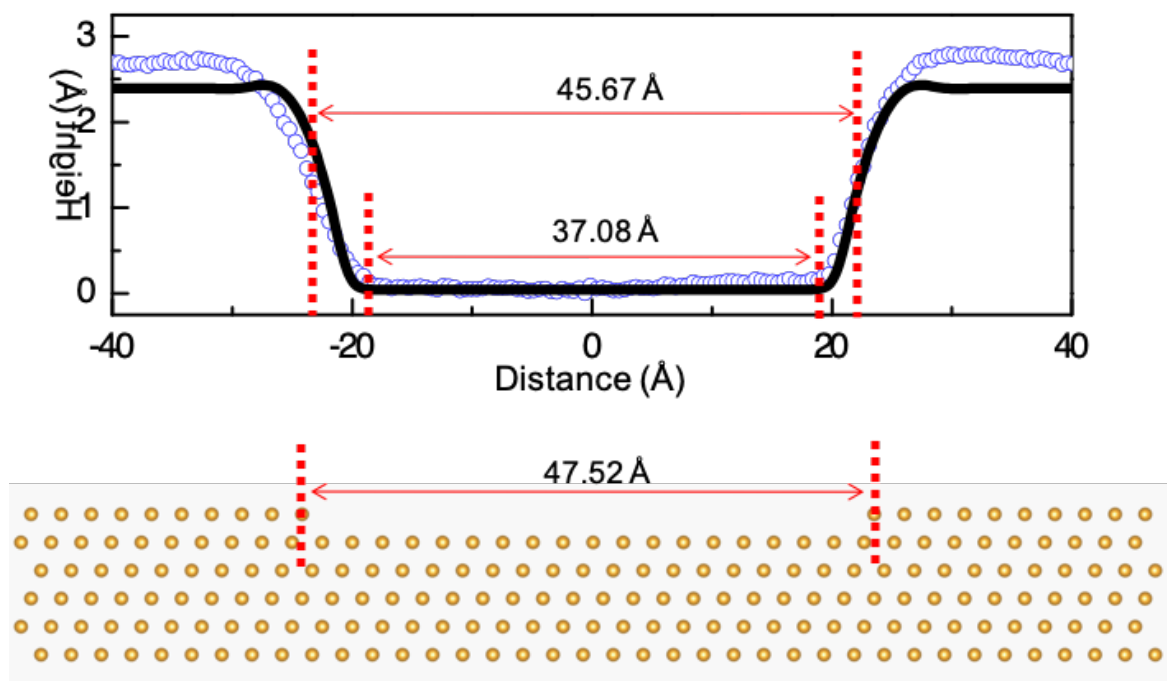

**Fig. S1.** Comparison of the experimental (blue open circles) and ESQC calculated (black line) constant current Δ*z* line scans across the Au(111) monoatomic step edges nano-indented trench. The experimental line scan is from Fig. 2c of main manuscript and was recorded at 100 mV/100 pA. The calculated scan was obtained with 10 mV/100 pA conditions in ESQC with a pyramidal 5-layer [111] facet STM tip apex atomic structure. A lateral atomic view of the trench is also presented at the bottom of the figure confirming the number of Au atoms (19) in the width of the trench imaged in Fig. 2 of the main manuscript.

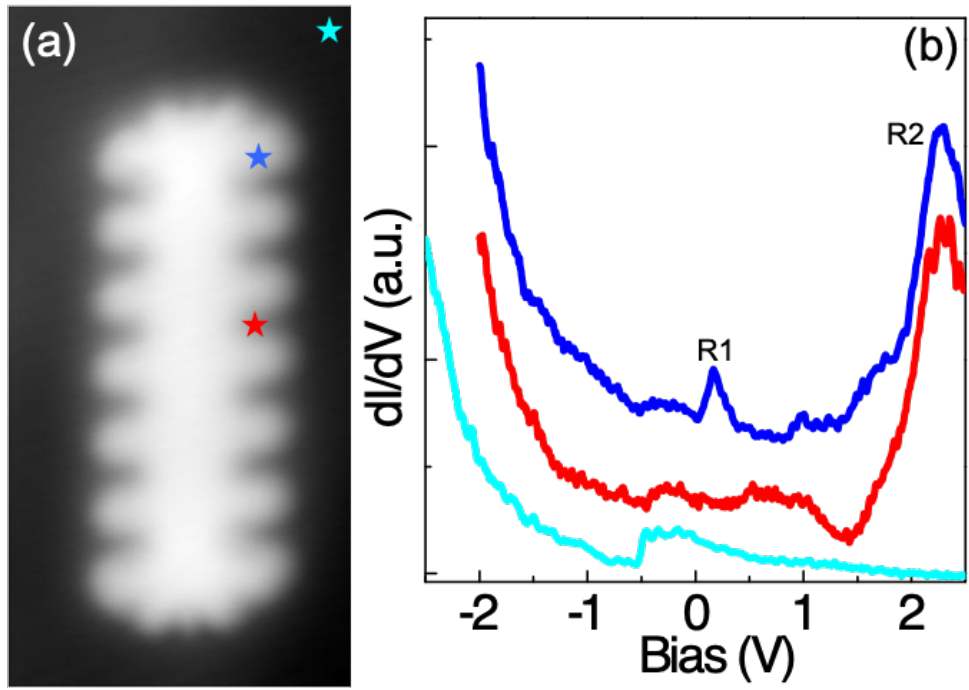

**Fig. S2. STM and STS of a 7-DP-GNR.** (a) The experimental constant current STM image of a planar 7-DP-GNR (scan area is 4 nm × 8 nm and scan parameters are 100 mV/100 pA) and (b) the d$I$/d$V$ spectra recorded at the edge (blue line) and center (red line) positions of the 7-DP-GNR revealing the characteristic tunneling electron resonances at +180 mV (R1) and +2.2 V (R2), respectively. We have selected the on-set positions of these two resonances located at ~+100 mV and ~+2.0 V for recording the d$I$/d$V$ maps that are presented in Figure 3 of the main manuscript. A reference Au(111) d$I$/d$V$ spectrum (cyan line) also presented in the figure for comparison.

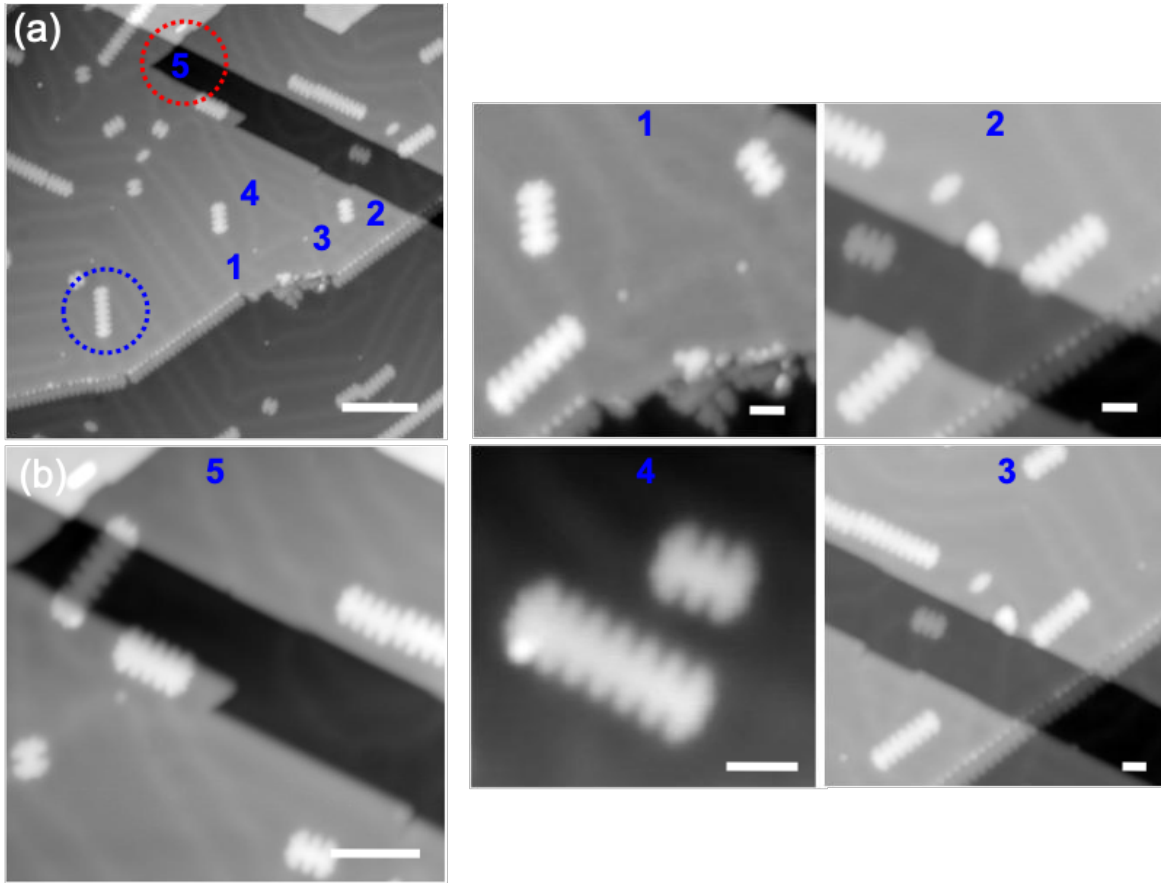

**Fig. S3. Manipulation of 7-GNR into bridged conformation.** (a) Large scale STM topograph showing the linear monoatomic high Au(111) trench (marked with a red dotted circle) and a 7-DP-GNR (marked with a blue dotted circle) compatible for bridging experiments. Corresponding step-by-step manipulation path of GNR to the linear Au(111) trench is indicated with the manipulation step number. The respective STM image recorded at each manipulation step is presented in the right panel images 1 to 4. (b) STM image showing the final 5th step after manipulating the 7-DP-GNR into a bridged conformation above the linear Au(111) trench. Scale bars are 10 nm in (a), 5 nm in (b) and 2 nm for STM images presented in the right panel. STM set-parameters are +100 mV/100 pA.

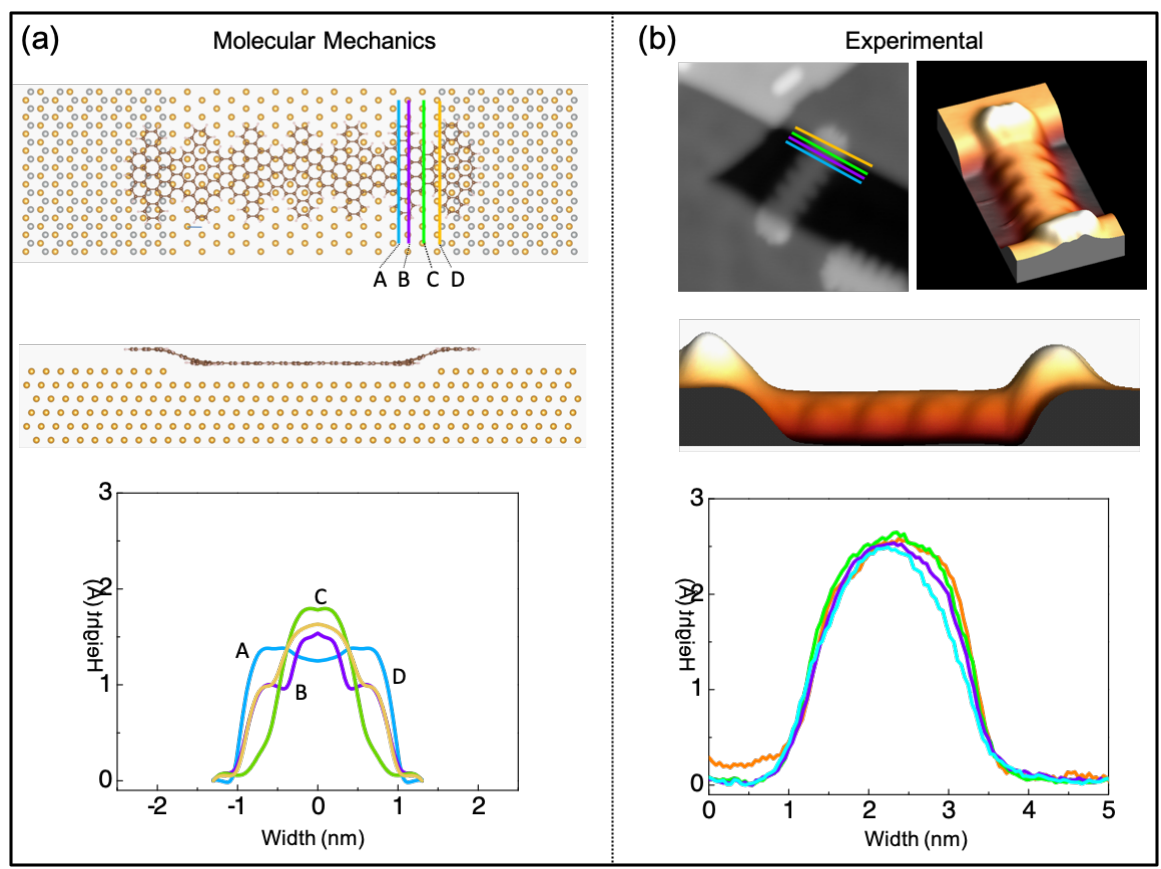

**Fig. S4.** ESQC Calculated (a) and corresponding experimental (b) Δ*z* constant current line scans transverse to the 7-DP-GNR at the different scan locations indicated with A, B, C, and D top view line on the 7-DP-GNR bridged conformation (top and lateral views of atomic model are also presented in the top left). The ESQC calculated scans were obtained with a low voltage +10 mV/ 100 pA conditions. Experimental line scans were recorded with 100 mV/100 pA. Top and lateral experimental 3D views are presented top right showing nicely the central deformation of the 7-DP-GNR. The ESQC calculated scans A, B, C and D are better resolved laterally than their experimental equivalents presented in (b). This is due to: (1) the perfect atomic structure of the tip apex in ESQC and (2) because ESQC is a mono-electronic calculation technique which is based on a description of the 7-DP-GNR molecular structure using a molecular orbitals description and not a full CI Slater like determinant superposition. Therefore, the ESQC mono-electronic calculations enhance the contribution of the corresponding molecular orbital (at 100 mV, the HOMO) as compared to the complete superposition of many molecular orbitals in a multi-electronic description.

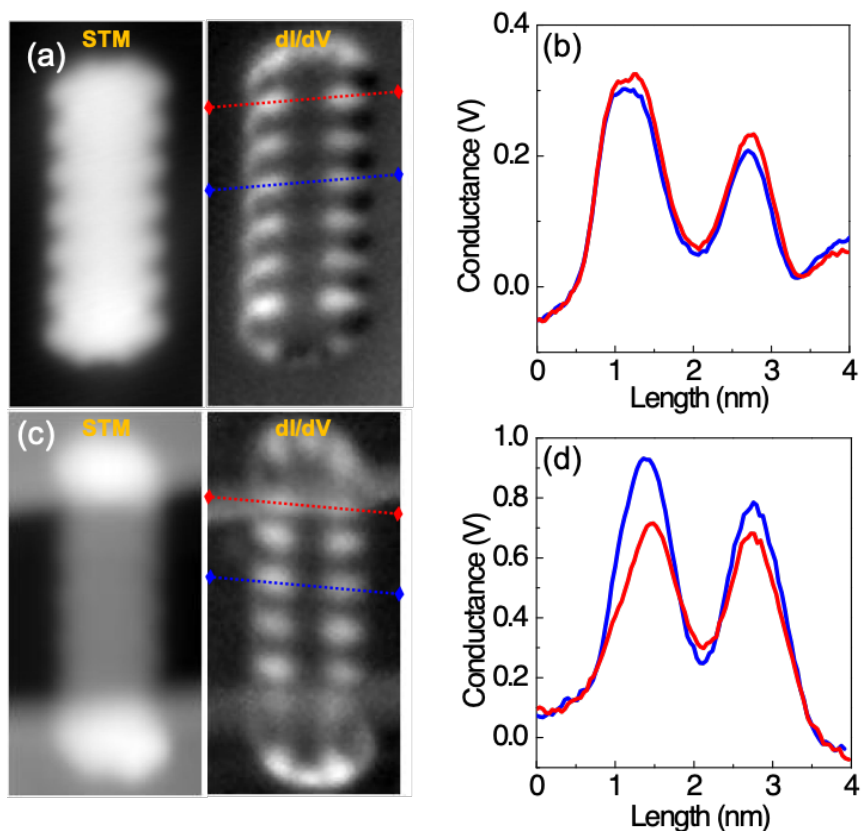

**Fig. S5.** The STM image (left) and the d$I$/d$V$ map (right) of the 7-DP-GNR presented in main manuscript (a) before and (c) after bridged conformation recorded at +2 V/100 pA. The scan sizes are 4 nm × 8 nm. Corresponding differential conductance line profiles (across the dotted lines in STM images) extracted from the d$I$/d$V$ maps at selected locations (b) before and (d) after bridged conformation indicate considerable impact of 7-DP-GNR bending conformation on the local d$I$/d$V$ features. Off-set in the line profiles along $y$-scale are adjusted for clarity.

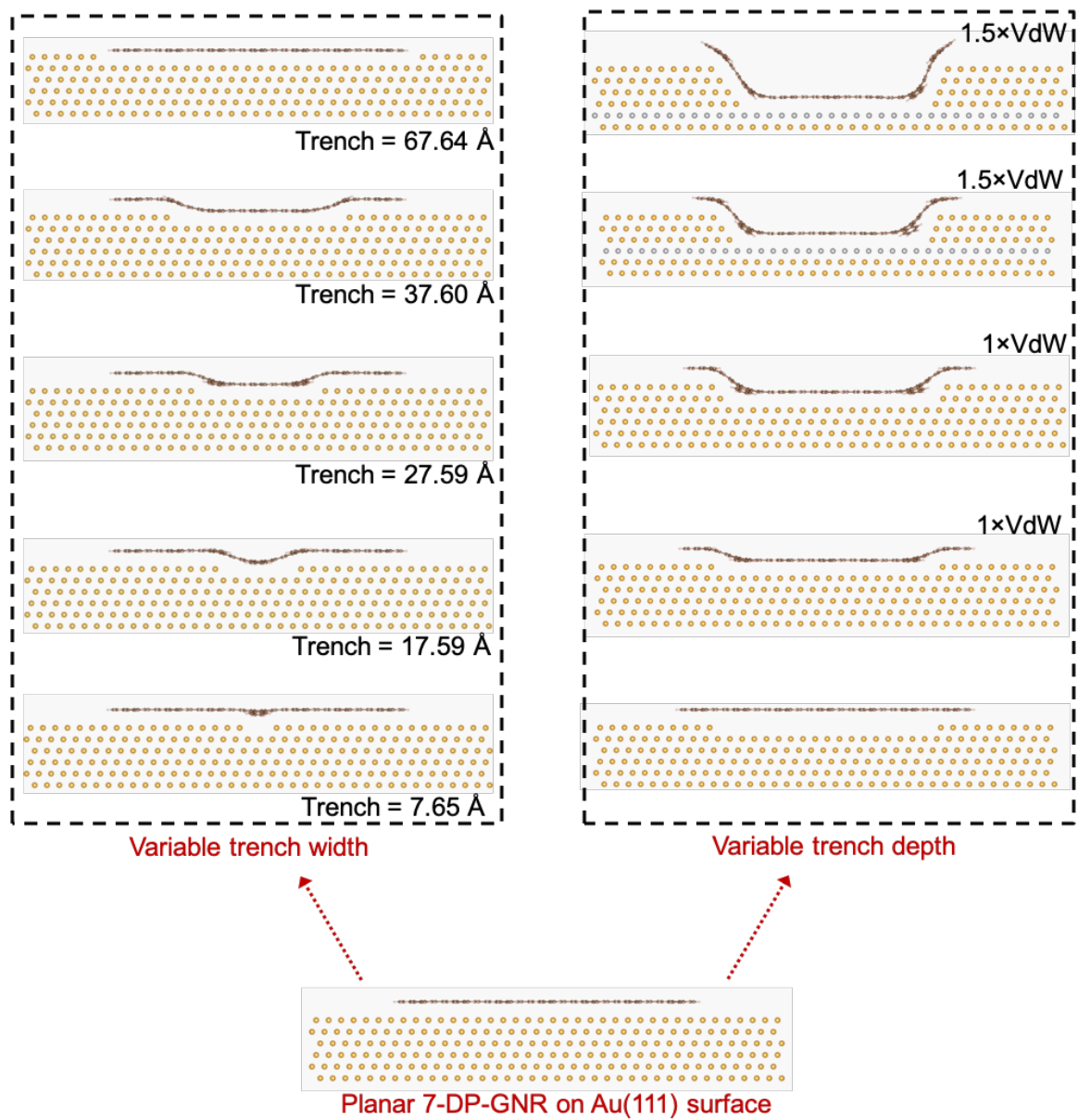

**Fig. S6.** Molecular mechanics ASED+ optimized conformation of a 7-DP-GNR positioned over the Au(111) trench by varying the trench width (left panel) and trench depth (right panel) starting from its initial planar conformation (bottom).

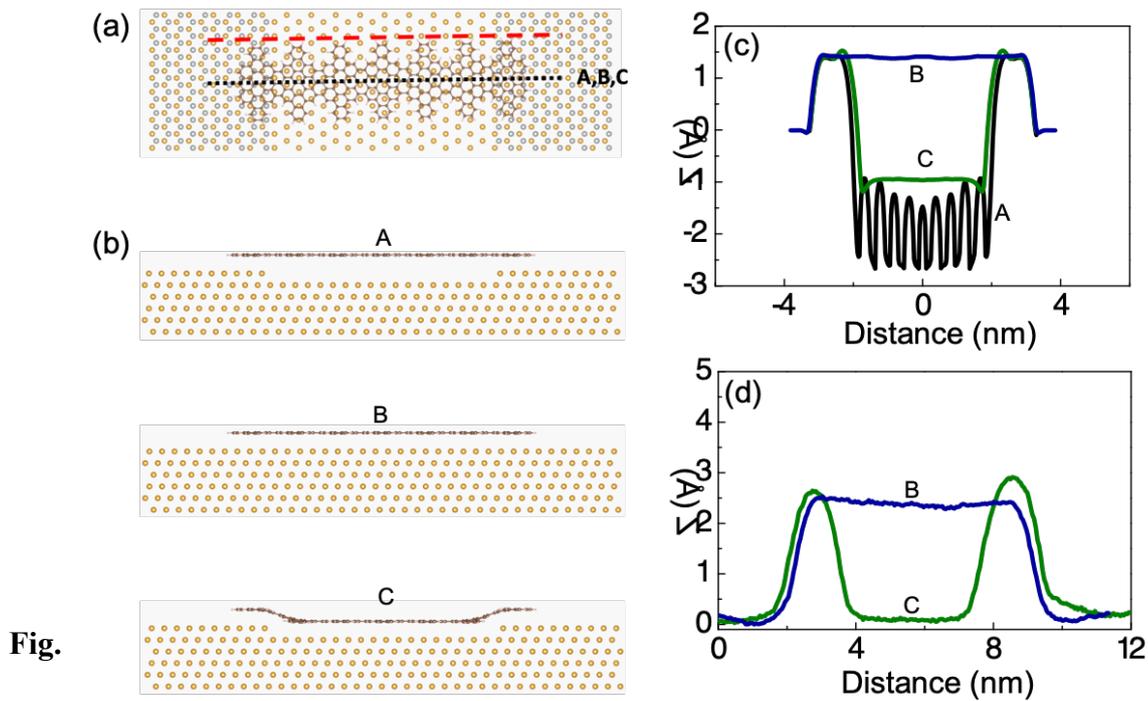

**S7.** ESQC calculated Δ$z$ constant current line scans longitudinal along the 7-DP-GNR molecular wire following the dotted black line in (a). (b) The 3 molecular configurations used for calculating the (c) line scans with (A) an artificial flat 7-DP-GNR conformation bridging the trench. (B) is the planar optimized standard conformation on the Au(111) fcc and (C) down-the-trench optimized conformation. As noticed in Fig. S4, the ESQC mono-electronic approximation enhances the molecular orbital contribution in calculated constant current line scan as evidenced for the calculated scan line (A) in (c) where the Δ$z$ oscillations are coming from the number of phenyl along the scan. For comparison (d) is reproducing the experimental scans extracted from Fig. 3e and 3f of the main manuscript. The comparison of the calculated and experimental blue scan variations (B) are satisfactory. There is a Δ$z$ discrepancy between the calculated and experimental green scans for (C). ESQC constant current scans were calculated with low voltages at 10 mV/100 pA conditions. The red dotted line in (a) is to indicate the line scan for the Fig. 5b of main manuscript.